\documentclass[aps,preprint,tighten,floats,epsf,rotate]{revtex4}
\usepackage{array}
\usepackage{graphicx}
\usepackage{float}

\begin{document}
\title {Effects of Phase Transition induced density fluctuations 
on pulsar dynamics}
\author{Partha Bagchi$^1$ \footnote {email: partha@iopb.res.in}}
\author{Arpan Das$^1$ \footnote {email: arpan@iopb.res.in}}
\author{Biswanath Layek$^2$ \footnote{email:
layek@pilani.bits-pilani.ac.in}}
\author{Ajit M. Srivastava$^1$ \footnote{email: ajit@iopb.res.in}}
\affiliation{$^1$ Institute of Physics, Sachivalaya Marg,
Bhubaneswar 751005, India, \\ $^2$ Department of Physics, Birla Institute
of Technology and Science, Pilani - 333031, India}

\begin{abstract}
\noindent 
We show that density fluctuations during phase transitions
in pulsar cores may have non-trivial effects on pulsar timings, and may also 
possibly account for glitches and anti-glitches. These density fluctuations
invariably lead to non-zero off-diagonal components of the moment of
inertia, leading to transient wobbling of star. Thus, accurate measurements 
of pulsar timing and intensity modulations (from wobbling) may be used to 
identify the specific pattern of density fluctuations, hence the 
particular phase transition, occurring inside the pulsar core. Changes in 
quadrupole moment from rapidly evolving density fluctuations during
the transition, with very short time scales, may provide a new source for 
gravitational waves.\\

\end{abstract}
\maketitle

\section{introduction}

\noindent 
The core of an astrophysical compact dense object, such as a neutron star, 
provides physical conditions where transition to exotic phases of 
quantum chromo dynamics (QCD) \cite{raja} may be possible. Superfluid phases of 
nucleons are also believed to exist inside neutron stars with vortex 
depinning associated with glitches (though it may not provide a viable 
explanation for anti-glitches). In this paper we propose a technique to 
probe the dynamical phenomena happening inside the neutron star, which 
may also account for glitches and anti glitches in a unified
framework.

We consider density fluctuations which invariably arise in any phase 
transition. We show that even with relatively very small magnitudes, 
these density fluctuations may be observable with accurate measurement 
of pulsar timings which can detect very minute changes in the moment
of inertia (MI) of the pulsar. This provides a very 
sensitive probe for density changes and density fluctuations
(especially due to formation of topological defects)
during phase transitions in the pulsar core. 
Non-zero off-diagonal components of moment of inertia arising from density
fluctuations imply that a spinning neutron star will develop wobble
leading to modulation of the peak intensity of pulses (as the direction
of the beam pointing towards earth undergoes additional modulation).
This is a unique, falsifiable, prediction of our model, that
rapid changes in pulsar timings should, most often, be associated with
modulations in changes in peak pulse intensity. It is important to
note that the vortex de-pinning model of glitches is not expected to
lead to additional wobble as the change in rotation caused by de-pinning
of vortex clusters remains along the rotation axis. Density fluctuations will
also lead to development of rapidly changing quadrupole moment which can
provide a new source for gravitational wave emission due to extremely
short time scales involved (despite small 
magnitude of this new contribution to the quadrupole moment) .

The effect of phase changes on the moment of inertia has been discussed
in literature. For example, moment of inertia change arising
from a phase change to high density QCD phase (such as to the QGP phase) is
discussed in ref.\cite{michange}. In the scenario of ref.\cite{michange},
the transition is driven by slow decrease
in the rotation speed of the pulsar, leading to increasing central density
causing the transition as the central density becomes supercritical.
It is assumed that as the supercritical core grows in size (slowly,
over the time scale of millions of years), it continuously converts
to the high density QGP phase (even when the transition is of first order).
Due to very large time scale, the changes
in moment of inertia are not directly observable, but observations of
changes in the braking index may be possible.

 Our work differs from these earlier works primarily in our focus on 
the  {\it rapidly evolving density fluctuations} arising during 
the phase transitions. This has not been considered before as far as we 
are aware. (Though effects of density inhomogeneities in a rotating
neutron star have been discussed \cite{inhmrot}.) Density
fluctuations inevitably arise during phase transitions, e.g. during
a first order transition in the form of nucleated bubbles, and may
become very important in the critical regime during a continuous
transition. The density fluctuations arising from a phase transition
become especially prominent if the transition leads to formation of
topological defects. Extended topological defects can lead to strong
density  fluctuations which can last for a relatively long time (compared
to the phase transition time). It is obvious that such randomly arising
density fluctuations will affect the moment of inertia of the star
in important ways. Most importantly, it will lead to development
of transient non-zero off-diagonal components of the moment of inertia,
as well as transient quadrupole moment. Both of these will disappear after
the density fluctuations decay away and the transition to a uniform
new phase is complete. Net change in moment of inertia will have this
transient part as well as the final value due to change to the new phase.
It seems clear that this is precisely the pattern of a glitch or
anti-glitch where rapid change in pulsar rotation is seen which slowly
and {\it only partially} recovers to the original value. As we mentioned
above, transient change in quadrupole moment will be important for 
gravitation wave emission. It is important to realize that the net 
change in MI (as discussed previously, e.g. in \cite{michange})
is only sensitive to the difference in the free energies of the two
phases, and cannot distinguish different types of phase transitions.
In contrast, density fluctuations arising during phase transitions
crucially depend on the nature of phase transition, especially on
the symmetry breaking pattern (e.g. via topological defects).
Identification of these density fluctuations via pulsar timings
(and gravitational waves) can pin down the specific transition
occurring inside the pulsar core. 

 We consider  the possibility of rapid phase transition
in a large core of a neutron star. The scenario of slow transition
as discussed in \cite{michange} is applicable for slowly evolving
star (e.g. by accretion), with a transition which is either
a weak first order, or a second order (or a crossover). If the
transition is strongly first order then strong supercooling can
lead to extremely suppressed nucleation rate, with transition
occurring suddenly after the supercritical core becomes macroscopic
in size. (This is similar to the situation of low nucleation
rate leading to nucleation distance scales of the order of meters in the
early universe as discussed by Witten \cite{wtn}.) Rapid transition
in a large core of neutron star is naturally expected during
the early stages of evolution of neutron star due to rapid cooling
of star. It could also be driven by rapid accretion on a neutron
star, coupled with a first order transition. 

  We will not discuss any detailed scenario of such a rapid transition
in this paper, and rather just note the possibility of such transitions.
Our focus will be that once such a rapid transition occurs, what are its
observational implications. Clearly, change in phase will lead to
net change in MI of star, affecting its rotation. This has been
discussed earlier, e.g. in \cite{michange}. We will focus on
additional effects associated with presence of density
fluctuations which lead to qualitatively new effects, not included
in the earlier investigations. These effects are, a transient
component of change in MI, development of the off-diagonal 
components of MI, and a transient, rapidly evolving quadrupole 
moment.

\section{Effects of density fluctuations due to bubble nucleation}

A rough estimate of change in the moment of inertia due to phase change
can be taken from ref.\cite{michange} using Newtonian approximation, and
with the approximation of two density structure of the pulsar. If the density
of the star changes from $\rho_1$ to a higher density $\rho_2$ inside
a core of radius $R_0$, then the fractional change in the moment of
inertia is of order,

\begin{equation}
{\Delta I \over I} \simeq {5 \over 3} 
({\rho_2 \over \rho_1} - 1){R_0^3 \over R^3}
\end{equation}

Here $R$ is the radius of the star in
the absence of the dense core. For a QCD phase transition, density changes
can be of order one. We take the density change to be about 30\% as an
example. If we take the largest rapid fractional change in the moment  of
inertia of neutron stars, observed so far (from glitches), to be less than
$10^{-5}$, then Eq.(1) implies that $R_0 \le$ 0.3 Km (taking $R$
to be 10 Km). For a superfluid transition, we may take
change in density to be of order of superfluid condensation energy
density $\simeq$ 0.1 MeV/$fm^3$ (see, ref.\cite{superfluid}).
In such a case, $R_0$ may be as large as 5 Km. These constraints on
$R_0$ arise from observed data on glitches/anti-glitches. These estimates
may also be taken as prediction of possible large fractional changes in
the moment of inertia (hence pulsar spinning rate) of order few percent
when a larger core undergoes rapid phase transition. For example, $R_0$
may be of order 2-3 km for QCD transition (from estimates of high density
core of neutron star \cite{nstar}), or it may be only slightly smaller
than $R$ for superfluid transition.

 We now discuss density fluctuations during phase transitions. First
we consider fluctuations arising simply from a first order transition.
Consider nucleation of relatively large number of bubbles (several
thousand) inside the supercritical core. This is
possible due to nonuniformities, even of purely statistical
origin (e.g. from fluctuations in temperature \cite{landau}).

\subsection {Parameters for bubble nucleation and Results }

We simulate random nucleation of bubbles filling up a core of size 
$R_0$ (= 300 meters). Effects of bubble nucleations will be characterized 
in terms of the following parameters.

Bubble radius $r_0$ will be taken to vary from 5 meters to 20 meters, 
with bubble separation being of same order as bubble size (close packing).

Density change in bubble nucleation is taken to be about the nuclear 
saturation density $\simeq$ 160 MeV/$fm^3$ (e.g. for QCD scale).

We find that density fluctuations lead to fractional change in MI, 
$\Delta I/I \simeq 4 \times 10^{-8}$ for $r_0 = 20$ meters implying similar 
changes in pulsar timings. Change in MI remains of same order when 
$r_0$ is changed from 20 meter to 5 meter. 
Due to random nature of bubble nucleation, off-diagonal
components of the moment of inertia, as well as the quadrupole moment
become nonzero and the ratio of both to the initial moment of inertia
are found to be of order $I_{xy}/I_0 \simeq Q/I_0 \simeq 10^{-11} - 10^{-10}$.
 
This aspect of our model is extremely
important, arising entirely due to density fluctuations generated during
the transition. As these density fluctuations homogenize, finally leading
to a uniform new phase of the core, both these components will dissipate
away. The off-diagonal component of moment of inertia will necessarily
lead to wobbling (on top of any present initially), which will get
restored once the density fluctuations die away. This will lead to
transient change not only in the pulse timing, but also in the pulse
intensity (as the angle at which the beam points towards earth gets
affected due to wobbling).  We again emphasize that
the conventional vortex de-pinning model of glitches is not expected to
lead to additional wobble as the change in rotation caused by de-pinning
of vortex clusters remains along the rotation axis. Thus, the presence of
intensity modulations associated with a glitch can distinguish between
our model and the vortex de-pinning model.

 Generation of quadrupole moment has obvious implication for gravitational
wave generation. One may think that a quadrupole moment of order $Q/I_0
\simeq 10^{-10}$
is too small for any significant gravity wave emission. However,
note that the gravitation power depends on the (square of) third time
derivative of the quadrupole moment \cite{gwave}. The time scales
will be extremely short here compared to the time scales considered in
literature for the usual mechanisms of change in quadrupole moment
of the neutron star. Here, phase transition dynamics will lead to
changes in density fluctuations occurring in time scales of microseconds
(or even shorter as we will see below in discussions of topological defects
generated density fluctuations). This may more than compensate for
the small amplitude of quadrupole moment and may lead to these density
fluctuations as an important source of gravitational wave emission from
neutron stars, as we will discuss below.

\section{Density fluctuations from topological defects}

Topological defects form during spontaneous 
symmetry breaking transitions via the so called 
{\it Kibble mechanism} \cite{kbl}. These defects can be source of
large density fluctuations depending on the relevant energy scales, and
their formation and evolution shows universal characteristics, (e.g. 
scaling behavior). This may lead to reasonably
model independent predictions for changes in MI, and quadrupole moment 
and subsequent relaxation. As bubbles, strings, domain walls, all 
generate different density fluctuation, with specific evolution patterns, 
high precision measurements of pulsar timings and intensity modulations 
(from wobbling) and its relaxation may be used to identify different 
sources of fluctuations, thereby pinning down the specific phase transition 
occurring. Specific phase transitions expected inside pulsars lead to
different types of topological defects. Important thing is that a
random network of defects will arise in any phase transition, and
resulting defect distribution can be determined entirely using the
symmetry breaking pattern. For example, superfluid transition
leads to formation of random network of vortices.
Confinement-deconfinement QCD transition can lead 
to formation of a network of domain walls and global strings arising 
from the spontaneous breaking of Z(3) center symmetry \cite{ranjita}. 
QCD transition may also give rise to only global strings, e.g.
in the color flavor locked (CFL) phase with $SU(3)_c \times 
SU(3)_L \times SU(3)_R \times U(1)_B$ symmetry (for 3 massless flavors)
is broken down to the diagonal subgroup $SU(3)_{c+L+R} \times Z_2$
\cite{raja}. As the resulting density fluctuations are mostly dominated
by the nature of defect (domain walls, or strings, or both), we will
model formation of these different defect networks in the following
using correct energy scale and try to estimate resulting density fluctuations.

\subsection{Results of model simulation of defect network}

 We model formation of U(1) strings and $Z_2$ domain walls 
by using  correlation domain formation
in a cubic lattice, with lattice spacing $\xi$ representing the correlation
length, as in ref. \cite{tanmay}. It is not possible to carry out these
simulation covering length scales of km (for star) to fm (QCD scale). Hence 
these simulations are necessarily restricted to very small system sizes.
Each lattice site is associated with an angle $\theta$ randomly
varying between 0 and 2$\pi$ (to model U(1) global string formation),
or two discrete values 0,1 (when modeling $Z_2$ domain wall formation).
For string case, winding of $\theta$ on each face of the cube is determined
using the geodesic rule. For a non-zero winding, a string segment (of
length equal to $\xi$ ) is assumed to pass through that
phase (normal to the phase). For domain wall case, any link connecting
two neighboring sites differing in $Z_2$ value is assumed to be intersected 
by a planar domain wall (of area $\xi^2$, and normal to the link). 
The mass density (i.e.  mass per unit length) of the string was taken as 
3 GeV/fm, and the domain wall tension is taken to be 7 GeV/fm$^2$.
These values are taken as order of magnitude estimates from the
numerical minimization results in ref. \cite{ranjita} for the pure 
gauge case. (Note that
logarithmic dependence of global strings on inter-string separation may 
lead to much larger density fluctuations than considered here.)

We consider spherical system of size $R$ and confine
defect network within a spherical core of radius $R_c = 
{0.3 \over 10} R$.  This is in view of the constraints on the supercritical
core size $R_c$ being of order 0.3 km for a neutron star with radius 
$R = 10$ km. Of course, in our simulations, $R$ is extremely small, with
maximum value of 4000 fm.
For $\xi \simeq $ 10 fm, we find the resulting value 
of  $\frac{\delta I}{I_i} \simeq 10^{-12} - 10^{-13}$ implying
similar changes in the rotational frequency. Here $I_i, i = 1,2,3$ are
the three diagonal values of the MI tensor. As we increase
$R_c$ from 5 $\xi$ to about 400 $\xi$, we find that the value of
$\frac{\delta I}{I_i}$ stabilizes near $ 10^{-13} - 10^{-14}$ as shown 
in the table below. This change in $\xi$ amounts to  change in the number 
of  string and wall segments by a factor of $10^6$. This 
gives a strong possibility that the same fractional change in the MI may 
also be possible when $R$ is taken to have the realistic value of about 
10 Km, especially when we account for statistical fluctuations in the
core.  For the formation of domain walls
we find fractional change in MI components (as well
as quadrupole moments) to be larger by about a factor of 40.
With accurate measurements, these small changes may be observable. 
Also, for a larger core size undergoing transition these changes can 
be larger.  We are only presenting
change in MI due to transient density fluctuations during phase
transition, which as we see, can have either sign. (As we discussed above,
the net change in MI
will include the very large contribution of order 10$^{-5}$ due to
net phase change of the core \cite{michange}.) This suggests that the
phase transition dynamics may be able to account for both glitch and
anti-glitch events (with associated wobbling from off-diagonal components
leading to pulse intensity modulation).

We now consider superfluid transition. A rapid superfluid transition 
could occur after transient heating of star (either due to another
transition releasing latent heat, or due to accretion etc). 
(In this work we neglect any possible effect of star rotation on this 
mechanism of random vortex formation which is expected to lead to a much
denser network than the one arising from star rotation.) We take the vortex
energy per unit length to be 100 MeV/fm and correlation length
for vortex formation of order 10 fm (ref.\cite{superfluid}). 
Table I shows that the string induced transient fractional change in MI 
is of order $10^{-10}$ (compared to net fractional change in MI of order 
$10^{-5}$ as discussed in Sect.II). Ratios of the quadrupole moment and 
off-diagonal components of MI to the net MI of the pulsar 
are also found to be of order $10^{-10}$. The transient change
in the MI decays away when the string system coarsens, thereby
restoring less than few percent of the original value. We note that this
pattern (and numbers) are similar to that of a glitch (or anti-glitch).

\begin{table}[h!]\footnotesize
\caption{Fractional change of various moments of the pulsar caused by 
inhomogeneities due to defects, with the correlation length $\xi =$ 10 fm.
For QCD scale strings, the string tension is taken as 3 GeV/fm, while 
the QCD Z(3) wall tension is taken as 7 GeV/fm$^2$ (from simulations in
ref.\cite{ranjita}). For the superfluid vortices, the energy per unit
length is taken to be 100 MeV/fm \cite{superfluid}.}
\vspace{0.25cm}
\begin{tabular}{|c|c|c|c|c|c|c|c|c|c|}
\hline
& \multicolumn{3}{|c|}{QCD Strings} & \multicolumn{3}{|c|}{QCD Walls} & \multicolumn{3}{|c|}{Superfluid Strings}\\
\hline
$\frac{R_c}{\xi}$ & $\frac{\delta I_{xx}}{I}$  & $\frac{\delta I_{xy}}{I}$
& $\frac{Q_{xx}}{I}$ &$\frac {\delta I_{xx}}{I}$ & $\frac {\delta I_{xy}}{I}$ & $\frac {Q_{xx}}{I}$
&$\frac {\delta I_{xx}}{I}$ & $\frac{\delta I_{xy}}{I}$ &$\frac {\delta Q_{xx}}{I}$\\
\hline 
5 & 5E-10 & -3E-10 & -1E-10 & 2E-8 & -1E-8 & -8E-10 &2E-6 &-1E-6 &-4E-7\\
\hline 
50 & 5E-12 & -2E-12 & 2E-12 &1E-10 & -8E-11 & -1E-11 &2E-8 &-7E-9 &7E-9\\
\hline
200 & 1E-13 & 2E-14 & -7E-14 & 5E-12 & -4E-12 & -6E-12 & 5E-10 & 6E-11 & -2E-10 \\
\hline
400 & -3E-15 & -5E-14 & -9E-14 & 3E-12 & -2E-12 & 3E-14 & -1E-11 & -2E-10 & -3E-10\\
\hline
\end{tabular}
\end{table}

\subsection{Field theory simulations  for QCD transition}

As a further support for these estimates, we have also carried out
field theory simulations of confinement-deconfinement (C-D) QCD
transition using effective field theory Polyakov loop model.
This leads to spontaneously broken Z(3) symmetry (for the SU(3) color 
group) in the QGP phase giving rise to topological domain wall 
defects in the QGP phase and also string defects forming at wall junctions.
We carry out a field theory simulation for the C-D
transition using a quench (quench is used for simplicity as only
domain formation is relevant here),
see ref.\cite{ranjita} for details.  It is not
possible to carry out field theory  simulation covering length scales of
km (for star) to fm (QCD scale). Hence these simulations are necessarily
restricted to system sizes of tens of fm only.
The physical size of the lattice  is taken as (7.5 fm)$^3$ and (15 fm)$^3$.
We use periodic boundary conditions and take
a spherical region with radius $R_c$ (representing star's core)
to study change of MI, with $R_c$ = 0.4 $\times$ lattice size.
We mention that the model of ref.\cite{ranjita} does not directly apply
to the case of neutron star which has large baryonic chemical potential. 
However, the most relevant features
of the model are formation of Z(3) defects with similar energy
scale as in the neutron star case. Hence we use those simulations
\cite{ranjita} to estimate defect induced density fluctuations
for the present neutron star case. We also use dissipation to relax 
density fluctuations but total energy is kept fixed by adding the 
dissipated energy.
Thus we only focus on re-distribution of energy in defect network and the
background. For the net change in the MI, we will use the 
estimates of $\Delta I/I$ from Sect.II (refs. \cite{michange}).
QCD transition may also give rise to only global strings, e.g.
in the color flavor locked (CFL) phase with $SU(3)_c \times
SU(3)_L \times SU(3)_R \times U(1)_B$ symmetry (for 3 massless flavors)
is broken down to the diagonal subgroup $SU(3)_{c+L+R} \times Z_2$
\cite{raja}. For this case, we modify the model studied in ref.
\cite{ranjita} by removing terms which correspond to Z(3) structure of
the vacuum manifold. This gives rise to string defects only without
any domain walls, with energy scale of QCD.

\begin{figure}[ht!]
\includegraphics[width=12cm]{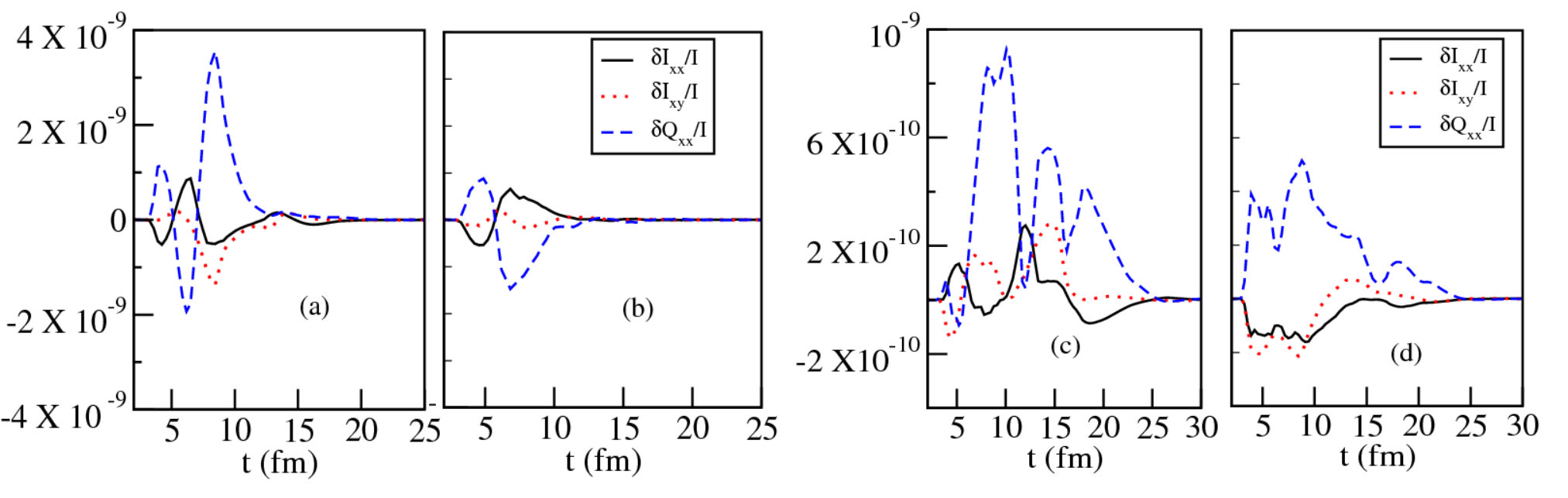}
\leavevmode
\caption{Fractional change in MI and quadrupole moment
during phase transitions. (a),(b) correspond to
lattice size (7.5 fm)$^3$, and (c),(d) correspond to lattice size
(15 fm)$^3$ respectively. Plots in (a) and (c) correspond to the
C-D phase transition with Z(3) walls and strings,
while plots in (b) and (d) correspond to the transition with
only string formation as for the CFL phase.}
\end{figure}
\vspace{0.25cm}

Plots in Fig.1 show the time evolution of the resulting fractional change
in the MI and the quadrupole moment of the core relative  to the initial 
total MI for these field theory simulations.
Here we have taken dense core of fractional size  0.3/10 (in view of
transition in core size of 300 meter for a 10 km star).
We add the MI of a shell outside the core so that total system size =
${10 \over 0.3}$ of the core size, and the shell has the same uniform
density as the core. Fractional changes for all components of MI as
well as the ratio of quadrupole moment to MI are found to be of similar
order. We are only presenting
change in MI due to transient density fluctuations during phase
transition, which as we see, can have either sign. (Net fractional change 
in MI will include the very large contribution of order 10$^{-5}$ due to
net phase change of the core \cite{michange}.) This suggests that the
phase transition dynamics may be able to account for both glitch and
anti-glitch events (with associated wobbling from off-diagonal components
leading to pulse intensity modulation). As discussed above, rapid
changes in the quadrupole moment will lead to gravitational wave emission.

\section{Gravitational wave generation due to density fluctuations}

Despite the small values of quadrupole moments in Table I, the power 
emitted in gravitational waves may not be small due to very short time 
scales in the present case. The defect coarsening will be governed by 
microphysics with time scale being of order tens of fm/c.
Even with extremely dissipative motion of strings, and much
larger length scales, the change in quadrupole moment due to strings
can happen in an extremely short time scale (during string
formation, and/or during string decay), thereby boosting the rate 
of quadrupole moment change. Even a very conservative value of the time 
scale of microseconds (e.g. for nucleation of bubbles 
with nucleation sites few meters apart, as discussed above) 
for the evolution of density fluctuations  will still be 1000 times 
smaller than fastest pulsar rotation time of  milliseconds,  
resulting in huge enhancement in the gravitational power,
as can be seen from the following expression for the power \cite{gwave}

\begin{equation}
{dE \over dt} = -{32G \over 5 c^5} {\Delta Q}^2 {\omega}^6 \simeq
- (10^{33} J/s) ({\Delta Q/I_0 \over 10^{-6}})^2 
({10^{-3} sec. \over \Delta t })^6
\end{equation}

Here $I_0$ is the MI of the pulsar, and 
$\Delta Q$ is the change in the quadrupole moment occurring in time 
interval $\Delta t$. For the present case, we can take $\Delta Q/I_0$ of 
order $10^{-14} - 10^{-10}$ from Table I (which is much smaller than the 
value of $10^{-6}$, typically used for deformed neutron stars). Even though
$\Delta Q/I_0$ is very small here, the relevant time scale $\Delta t$ is
also very small. Thus, even with a conservative estimate of the time scale 
for the transition (e.g. bubble coalescence), $\Delta t = 10^{-6} - 10^{-5}$ 
sec., the power in gravitational wave can be significant due to large
enhancement from the $({10^{-3} sec. \over 
\Delta t})^6$ factor.  A shorter time scale may make this source as 
potentially very prominent for  gravitational waves. For an estimate of 
the expected strain amplitude from a pulsar at a distance $r$, we use the 
following expression \cite{gwave}

\begin{equation}
h = {4\pi^2 G \Delta Q f^2 \over c^4 r} \simeq 10^{-24} ({\Delta Q/I_0
\over 10^{-6}}) ({10^{-3} sec. \over \Delta t })^2 ({1 kpc \over r})
\end{equation}

With $\Delta Q/I_0$ of order $10^{-10}$ and a time scale for
the transition (e.g. bubble coalescence), $\Delta t = 10^{-6} - 10^{-5}$ 
sec. we can have $h \simeq 10^{-24} - 10^{-22}$ for a pulsar at
1 kpc distance. As we mentioned above, the time scale for evolution of 
density fluctuations may be even shorter, leading to larger strain 
amplitudes (even if much smaller values of $\Delta Q/I_0$ are taken
from Table I), and much larger power emitted
in gravitational waves. Since the wave emission is only for a single
burst, lasting for only duration $\Delta t$, net energy lost by the star 
remains small fraction of the star mass. 

\section{conclusions}

 To summarize our results, we have shown that density fluctuations
arising during a rapid phase transition lead to transient change in 
the MI of the star. Such density fluctuations in general lead to
non-zero off-diagonal components of moment of inertia tensor which will
cause the wobbling of pulsar, thereby modulating the peak intensity of the
pulse.  This is a distinguishing and falsifiable signature of our model.
The conventional vortex de-pinning model of glitches
is not expected to lead to additional wobble as the change in rotation
caused by de-pinning of vortex clusters remains along the rotation axis.
We find that moment of inertia can increase or decrease, which gives
the possibility of accounting for the phenomenon of glitches and anti-glitches
in a unified framework. Development of nonzero value of quadrupole moment
(on a very short time scale) gives the possibility of gravitational
radiation from the star whose core is undergoing a phase transition.
Net change in MI (as discussed previously, e.g. in \cite{michange})
is only sensitive to the difference in the free energies of the two
phases, and cannot distinguish different types of phase transitions.
In contrast, density fluctuations arising during phase transitions
crucially depend on the nature of phase transition, especially on
the symmetry breaking pattern (e.g. via topological defects).
Identification of these density fluctuations via pulsar timings
(and gravitational waves) can pin down the specific transition
occurring inside the pulsar core. This is an entirely new way of 
probing phase transitions occurring inside the core of a neutron star.
Though our estimates suffer from the uncertainties of
huge extrapolation involved from the core sizes we are able to simulate,
to the realistic sizes, they strongly indicate that expected changes
in moment of inertia etc. may be well within the range of observations,
and in fact may be able to even account for the phenomena of
glitches and anti-glitches. 

  One aspect of glitches which may raise concern in our model is multiple
occurrences of glitches. For vortex de-pinning model multiple glitches
seem natural due to formation of vortices due to star rotation
with glitches occurring with de-pinning of clusters of vortices. In our
model, multiple occurrence of glitches will require multiple phase
transitions. This is not very improbable, though, when the transition
happens due to accretion from companion star causing heating, with 
subsequent cooling down. Essentially, in this case, the matter in the neutron 
star core will continue to remain in the vicinity of the phase boundary 
in the $T-\mu$ plane (in the QCD phase diagram). Accretion will move the 
system across the boundary by increasing $\mu$, causing a phase 
transition, and subsequent cooling will move the system back through 
the phase boundary with another transition. Such processes could in 
principle repeat for certain neutron stars leading to multiple glitches. 
One should also allow the possibility that glitches could occur due to 
multiple reasons. Some glitches/anti-glitches could occur due to the model
proposed here, that is due to phase transition induced density
fluctuations, while other glitches could occur due to the
conventional de-pinning of vortex clusters. Main point of our
work is to emphasize that if and when a transition happens in the
core of a neutron star, it invariably leads to density fluctuations
which manifest itself in glitch/anti-glitch like behavior, along with
other   implications such as wobbling of star, gravitational wave emission
etc.

It is thus important to investigate this interesting possibility
further.  With much larger simulations, and accounting
for statistical fluctuations of temperature, chemical potential etc. in
the core, more definitive patterns of changes in moment of inertia
tensor/quadrupole moment etc. may emerge which may carry unique
signatures of specific phase transitions involved. (For example,
continuous transitions will lead to critical density fluctuations, and
topological defects will induce characteristic density fluctuations depending
on the specific symmetry breaking pattern.) If that happens then this method
can provide a rich observational method of probing the physics of strongly
interacting matter in the naturally occurring laboratory, that is
interiors of neutron star. It will be interesting to see if any
other astrophysical body, such as white dwarf, can also be probed in a
similar manner.

\section{Acknowledgments}

 We thank A.P. Mishra, R.K. Mohapatra, A. Atreya, and S.S. Dave
for useful discussions. B.Layek would like to thank the HEP group at IOP for
hosiptality during his sabbatical leave from Physics Dept., BITS-Pilani, 
Pilani.

\end{document}